\documentclass[aps,prd,email,preprint,showpacs,showkeys,preprintnumbers,amsmath,amssymb,nofootinbib]{revtex4}

\usepackage{color}
\usepackage{latexsym}
\usepackage{amsmath}
\usepackage{amssymb}
\usepackage{eufrak}
\usepackage{euscript}
\usepackage{pstricks}
\usepackage{graphics}
\usepackage{graphicx}
\usepackage{picture}
\def\[{\left\lbrack}
\def\]{\right\rbrack}

\def\({\left(}
\def\){\right)}

\newcommand{\bee}{\begin{equation}}
\newcommand{\eee}{\end{equation}}
\newcommand{\eaa}{\end{eqnarray}}
\newcommand{\baa}{\begin{eqnarray}}

\def\ni{\noindent}

\begin{document}

\title{\Large{New bounds for Tsallis parameter in a noncommutative phase-space entropic gravity \\ and nonextensive Friedmann equations}}

\author{Everton M. C. Abreu$^{a,b}$}
\email{evertonabreu@ufrrj.br}
\author{Jorge Ananias Neto$^{b}$}
\email{jorge@fisica.ufjf.br}
\author{Albert C. R. Mendes$^b$}
\email{albert@fisica.ufjf.br}
\author{Wilson Oliveira$^b$}
\email{wilson@fisica.ufjf.br}

\bigskip

\affiliation{$^a$Grupo de F\' isica Te\'orica e Matem\'atica F\' isica, Departamento de F\'{\i}sica, \\
Universidade Federal Rural do Rio de Janeiro\\
BR 465-07, 23890-971, Serop\'edica, RJ, Brazil \\\\ 
${}^{b}$Departamento de F\'{\i}sica, ICE, Universidade Federal de Juiz de Fora,\\
36036-330, Juiz de Fora, MG, Brazil\\\\
\today\\}
\pacs{04.50.-h, 05.20.-y, 05.90.+m}


\begin{abstract}
\noindent In this paper, we have analyzed the nonextensive Tsallis statistical mechanics in the light of Verlinde's formalism.   We have obtained, with the aid of a noncommutative phase-space entropic gravity, a new bound for Tsallis nonextensive (NE) parameter (TNP) that is clearly different from the ones present in the current literature.  We derived the Friedmann equations in a NE scenario.  We also obtained here a relation between the gravitational constant and the TNP.
\end{abstract}


\maketitle

\pagestyle{myheadings}
\markright{New bounds for Tsallis parameter in a noncommutative...}





\section{Introduction}
\renewcommand{\theequation}{1.\arabic{equation}}
\setcounter{equation}{0}

There are theoretical evidences that the understanding of gravity has been greatly benefited from a possible connection with thermodynamics. Pioneering works of Bekenstein \cite{BEK} and Hawking \cite{HAW} have described this issue. For example, quantities as area and mass of black-holes are associated with entropy and temperature respectively. Working on this subject, Jacobson \cite{Jac} interpreted Einstein field equations as a thermodynamic identity. Padmanabhan \cite{PAD} gave an interpretation of gravity as an equipartition theorem. 

Recently, Verlinde \cite{verlinde} brought an heuristic derivation of gravity, both Newtonian and relativistic, at least for static spacetime.  The equipartition law of energy has also played an important role. 
On the other hand, one can ask: what is the contribution of gravitational models in thermodynamics theories?  

The concept introduced by Verlinde is analogous to Jacobson's \cite{Jac}, who proposed a thermodynamic derivation of Einstein's equations.  The result has shown that the gravitation law derived by Newton can be interpreted as an entropic force originated by perturbations in the information ``manifold" caused by the motion of a massive body when it moves 
away from the holographic screen.  Verlinde used this idea  together with the Unruh result \cite{unruh} and he obtained Newton's second law.   The entropic force together with the holographic principle and the equipartition law of energy resulted in Newton's law of gravitation.  Besides, Verlinde's ideas have been used in cosmology \cite{ma}.

There is an extension of the usual Boltzmann-Gibbs theory (BG) that is called Tsallis statistical theory \cite{tsallis} (TT). To sum up, the formalism initially considers the entropy formula as a nonextensive (NE) quantity where there is a parameter $q$ that measures the so-called degree of nonextensivity. This formalism has been successfully applied in many physical models.  An important feature is that when $q\rightarrow1$ we recover the usual Boltzmann-Gibbs theory, i.e., we have an extensive theory. 

The purpose of this paper is to use Verlinde's formalism in a noncommutative (NC) scenario \cite{bbdp,ion,ghosh} in order to determine a new bound for the NE parameter $q$.  The main motivation, in few words, is that noncommutativity and TT share a fractal geometry feature.  
We will see that when we compare the gravitational constant modified by NC effects with the gravitational constant modified by NE effects, a new and more precise bound for the NE parameter $q$ appears.

We will organize our paper in the following way: in section \ref{vf} we will provide a brief review of Verlinde's formalism. In section \ref{TTS} we will introduce the main steps of Tsallis statistical approach. In section \ref{Res} we will compute a new bound for the NE parameter $q$. In section \ref{FET} we derive the Friedmann equation in a NE scenario. The conclusions will be depicted in the last section.

\section{A Brief Review of Verlinde's Formalism}
\label{vf}
\renewcommand{\theequation}{2.\arabic{equation}}
\setcounter{equation}{0}

One of the reasons that the study of entropy has been an interesting task through the years is the fact that it can be considered as a measure of information loss concerning the microscopic degrees of freedom of a physical system when depicting it in terms of macroscopic variables.  Appearing in different scenarios, entropy can be deemed as a consequence of gravitational framework \cite{nicolini}.

The formalism proposed by E. Verlinde \cite{verlinde} derives the gravity acceleration  by using, basically, the holographic principle and the equipartition law of energy. His ideas relied on the fact that gravitation can be considered universal and independent of the details of the spacetime microstructure \cite{nicolini}.  Besides, Verlinde brought new concepts about holography since the holographic principle must unify matter, gravity and quantum mechanics \cite{rev}.

The model considers a spherical surface as being the holographic screen, with a particle of mass M positioned in its center. A holographic screen can be imagined as a storage device for information. The number of bits (the term bit means the smallest unit of information in the holographic screen) is assumed to be proportional to the area $A$ of the holographic screen
\begin{eqnarray}
\label{bits}
N = \frac{A }{l_p^2}\,\,,
\end{eqnarray}
where
\begin{equation}
\label{2.1a}
l_p = \sqrt{\frac{G\hbar}{c^3}}\,\,\,
\end{equation}
and $ A = 4 \pi r^2$  is the Planck length and $l_p^2$ is the Planck area.   In Verlinde's formalism we assume that the total energy of the bits on the screen is given by the equipartition law of energy

\begin{eqnarray}
\label{eq}
E = \frac{1}{2}\,N k_B T.
\end{eqnarray}
It is important to mention here that the usual equipartition theorem, Eq.(\ref{eq}), is derived from the usual Boltzmann-Gibbs statistics. We will see that in a NE statistical scenario, the equipartition law of energy will be modified in a way that a NE parameter $q$ will be introduced in its expression.
Considering that the energy of the particle inside the holographic screen is equally divided through all bits, we can write the equation

\begin{eqnarray}
\label{meq}
M c^2 = \frac{1}{2}\,N k_B T\,\,,
\end{eqnarray}

\ni where $M$, in the holographic principle is the mass that would appear in the region of space enclosed by the screen.  A test particle with mass $m$ can feel the energy distributed over the occupied bits.  It is equivalent to the emergent $M$ enclosed by the screen.  This can easily be seen since Eq. (\ref{eq}) represents the total energy.

An observer in the rest frame of this test mass, which is an accelerated frame, will evidence the existence of a temperature

\begin{eqnarray}
\label{un}
k_B T = \frac{1}{2\pi}\, \frac{\hbar a}{c}\,\,,
\end{eqnarray}

\ni due  to the Unruh effect \cite{unruh}, where $a$ is the acceleration of this frame.  Therefore, in the entropic theory, the Unruh temperature will be taken as the holographic screen temperature.  At this point, it is important to mention that, by using the equivalence principle, the acceleration $a$ given by the Unruh formula will also be the gravitational acceleration associated with a massive body in Verlinde's formalism.  

We are  in a position to derive the  (absolute) gravitational acceleration formula.  Substituting Eq. (2.1) in Eq. (2.4) and using Eq. (2.5) we can write that

\begin{eqnarray}
\label{acc}
a =  \frac{l_p^2 c^3}{\hbar} \, \frac{ M}{r^2}\nonumber\\ 
= G \, \frac{ M}{r^2}.
\end{eqnarray}
We can observe from Eq. (\ref{acc}) that the Newton constant $G$ is just written in terms of the fundamental constants, $G=l_p^2 c^3 / \hbar$.  










\section{The Tsallis statistical theory}
\label{TTS}
\renewcommand{\theequation}{3.\arabic{equation}}
\setcounter{equation}{0}

An important formulation of the NE Boltzmann-Gibbs statistical theory has been proposed by Tsallis  \cite{tsallis} in which 
$q$, known in the current literature as Tsallis parameter or NE parameter, is a real parameter which quantifies the degree of nonextensivity. 
The definition of entropy in TT motivates the study of multifractals systems and it also possesses the usual properties of positivity, equiprobability, concavity and irreversibility.
It is important to stress that Tsallis formalism contains the Boltzmann-Gibbs statistics as a particular case in the limit $ q \rightarrow 1$ where the usual additivity of entropy is recovered. Plastino and Lima  \cite{PL}, by using a generalized velocity distribution for free particles  \cite{SPL,MPP} 
have derived a NE equipartition law of energy whose expression can be written as

\begin{eqnarray}
\label{ge}
E = \frac{1}{5 - 3 q} N k_B T\,\,,
\end{eqnarray}
where the range of $q$ is $ 0 \le q < 5/3 $, for obvious reasons.  For $ q=5/3$ (critical value) the expression of the equipartition law of energy, Eq. (\ref{ge}), diverges. It is also easy to observe that for $ q = 1$,  the classical equipartition theorem for each microscopic degrees of freedom is recovered. 

As another path, it is worth to mention that in \cite{cn} the authors obtained that Tsallis entropy can be written as
\bee
\label{AAA}
S_q\,=\,k_B\,\frac{\Gamma^{q-1}\,-\,1}{q\,-\,1}\,\,,
\eee
where $\Gamma$ is proportional to the volume, $\Gamma \equiv V$.  Let us consider that the linear size is given by $L$, so $\Gamma \equiv V\,=\,L^3$.   Substituting this value into the $\Gamma$ factor of Eq. (\ref{AAA}) we have that

\bee
\Gamma^{q-1}\,=\,L^{3(q-1)}\,\,.
\eee

\ni Then, it is easy to see that for $q=5/3$ we have that $\Gamma^{q-1}\,=\,L^2 \rightarrow S_q \propto L^2\,=\,A$ and we can say that 	Eq. (\ref{AAA}) scales as the area, which is just the expression for the black hole area entropy law.   The consequence is that, due to the nonextensivity of entropy in Eq. (\ref{AAA}), a holographic screen, which is underlying in Verlinde's framework, is originated.   


As an application of NE equipartition theorem in Verlinde's formalism we can
substitute the equipartition law by the NE equipartition formula, i.e., Eq. (\ref{ge}) into  Eq. (\ref{meq}).  By applying the same steps described in section \ref{vf}, we can obtain a modified acceleration formula given by

\begin{eqnarray}
\label{accm}
a = G_{NE} \, \frac{ M}{r^2},
\end{eqnarray}
where $G_{NE}$ is an effective gravitational constant which is written as

\bee
\label{S}
G_{NE}=\,\frac{5-3q}{2}\,G\,\,.
\eee
From result (\ref{S}) we can observe that the effective gravitational constant depends on the NE parameter $q$. For example, $q=1$ we have $ G_{NE}=G$ (BG scenario) and for $q\,=\,5 / 3$ we have the curious and hypothetical result which is $G_{NE}=0$.  This result shows us that $q\,=\,5/3$ is an upper bound limit when we are dealing with the holographic screen.  Notice that this approach is different from the one demonstrated in \cite{cn}, where the authors considered in their model that the number of states is proportional to the volume and not to the area of the holographic screen.

These last considerations motivates us to present a preliminary calculation of the $q$-bound.
From Eq. (\ref{S}), it is possible to obtain a bound for the Tsallis parameter from the experimental values of the gravitational constant.  For example, recent results from CODATA \cite{codata} show that
\begin{eqnarray}
\label{GC}G=6.673 84(80) \times 10^{-11} m^3 kg^{-1} s^{-2},
\end{eqnarray}
with a relative standard uncertainty of $1.2 \times 10^{-4}$\,\,.  From Eq. (\ref{S}) we can write that
\bee
\label{SS}
1-q\,=\,\frac 23 \,\frac{G_{NE}-G}{G}\,\,.
\eee
If we assume that the right hand side of Eq. (\ref{SS}) is the uncertainty, then we have
\begin{eqnarray}
\label{1B}
|q-1|\leq {\cal{O}}\cdot 10^{-4}\,\,.
\end{eqnarray}
The bound above is of the same order as those obtained in \cite{TBL,TT}.  

As we said before, the $|q-1|\sim 0$ limit takes us back to Boltzmann entropy.  But let us clarify a little more about this issue.  Let us begin analyzing the generalized equipartition theorem given in Eq. (\ref{ge}) embedded in Verlinde's scenario, where the bits now obey a NE statistical mechanics.  At this point it is clear that Verlinde's formalism allows us to consider changes in the physical properties of specific gravitational system when the equipartition law of energy is modified under determined rules.  Let us assume that the number of bits is

\bee
\label{2a}
N\,=\,\frac{A}{l^2}\,\,,
\eee

\ni where $A$ is the same as in (\ref{bits}) and it is equal to $4\pi r^2$ and $l$ is a fundamental length different from $l_p$.  So, let us consider, as before, that $E=Mc^2$, localized at the center of the holographic screen.  If we combine Eqs. (\ref{un}), (\ref{ge}) and (\ref{2a}) we can write that

\baa
\label{2b}
a\,&=&\,\frac{5-3q}{2}\,\frac{l^2 c^3}{\hbar}\,\frac{M}{r^2} \nonumber \\
&\Longrightarrow& a\,=\,G\,\frac{M}{r^2}\,\,,
\eaa

\ni where the Newton constant $G$ is given by

\bee
\label{2c}
G\,=\,\frac{5-3q}{2}\,\frac{l^2 c^3}{\hbar}
\eee

\ni and from (\ref{2.1a}) we have that

\bee
\label{2d}
l^2_p\,=\,\frac{5-3q}{2}\,l^2
\eee

\ni and for $q=1$ we have that $l_p^2\,=\,l^2$ (more details in \cite{ananias1}).  From (\ref{2d}) we see that the fundamental length depends on the $q$-parameter, i.e., the nonextensive parameter.  In the future we will talk again about the limit $q \rightarrow\; 1$. We will see that Tsallis $q$-extension of statistics and NC theories are two faces of the same physical reality which is the fractal geometry of spacetime.

\section{A New precise Bound for the Nonextensivity Parameter through Noncommutativity}
\renewcommand{\theequation}{4.\arabic{equation}}
\setcounter{equation}{0}
\label{Res}

As we saw in the last sections, entropy promotes the connection between gravity and the microstructure of a quantum spacetime.  It is very well established that to understand NC quantum theory can lead us to realize the details of the very early Universe physics  \cite{reviews} and of the quantum spacetime where the concept of a discrete spacetime is believed to be substituted by the one of fuzziness (fractal).  

Noncommutativity was rekindled when it was also described in superstring/M-theory.  In few words we can explain that a NC algebra arises when describing the excitations of open strings in the presence of a Neveau-Schwarz constant background field  \cite{sw}.  We can also see noncommutativity in other areas of research such as entropic gravity  \cite{nicolini,bbdp}, quantum cosmology  \cite{bbdp2}, the Schwarszchild black hole thermodynamics  \cite{bbdp3}, the black hole singularity  \cite{bbdp4} and wormhole physics \cite{wormhole}.
In this section we will make use of the results obtained in  \cite{bbdp} to obtain a new bound for the NE parameter using NC space.

The geometry that governs a NC space is based on a canonical phase-space NC algebra.  Let us describe our $d$-dimensional phase-space such as  \cite{bbdp}
\bee
\label{B0}
[\hat{q}_i , \hat{q}_j]\,=\,i\theta_{ij}\quad,\:[\hat{q}_i , \hat{p}_j]\,=\,i\hbar\,\delta_{ij}\quad,\: 
[\hat{p}_i , \hat{p}_j]\,=\,i\eta_{ij}\,\,\qquad i,j\,=\,1,\ldots,d\,\,,
\eee

\ni where $\eta_{ij}$ and $\theta_{ij}$ are antisymmetric real constant $d\times d$ matrices and $\delta_{ij}$ is the identity matrix.

The extended algebra in (\ref{B0}) is related to the basic well known Heisenberg-Weyl algebra.  The mapping between commutative and NC variables is named mathematically as Darboux maps (the Seiberg-Witten map, in physics) which are not unique.  However, all physical quantities like expectation values, eigenvalues and probabilities are independent of the specific chosen Darboux map (\cite{bbdp}, and references therein).

Now we will describe the NC correction to the gravitational force developed in  \cite{bbdp}.  The basic idea is that the Planck constant is corrected by NC effects. We begin by considering that a two dimensional phase-space cell has a minimal volume $\hbar^2$ given by

\bee
\label{B12}
V(\Delta x_1,\Delta p_1,\Delta x_2,\Delta p_2)\,=\,\Delta x_1\,\Delta p_1\,\Delta x_2\,\Delta p_2\,\,,
\eee

\ni where $\Delta x_1\,\Delta p_1\,\geq \frac{\hbar}{2},\,\Delta x_2\,\Delta p_2\,\geq \frac{\hbar}{2},\,
\Delta x_1\,\Delta x_2\,\geq \frac{\theta_{12}}{2}$ and 
$\Delta p_1\,\Delta p_2\,\geq \frac{\eta_{12}}{2}$ are the constraints.  For more details see  \cite{bbdp}.  If we want to minimize the product $\Delta x_1\,\Delta x_2\,= \frac{\theta_{12}}{2}$ we have to use that 
$\Delta p_1\,\Delta p_2\,= \frac{\eta_{12}}{2}$, hence,

\bee
\label{B13}
V(\Delta x_1,\Delta p_1,\Delta x_2,\Delta p_2)\,\geq\,\frac{\theta_{12}\eta_{12}}{4}\,\,.
\eee

\ni To minimize $\Delta x_1 \Delta p_1$ we have to use that $\Delta x_1 \Delta p_1=\frac \hbar2\,=\Delta x_2 \Delta p_2$, so we can write that

\bee
\label{B14}
V(\Delta x_1,\Delta p_1,\Delta x_2,\Delta p_2)\,\geq\,\frac{\hbar^2}{4}
\eee

\ni and the total volume is 

\bee
\label{B14prime}
V(\Delta x_1,\Delta p_1,\Delta x_2,\Delta p_2)\,\geq\,\frac{\hbar^2}{4}\,+\,\frac{\theta_{12}\eta_{12}}{4}
\,=\,\frac{\hbar^2_{eff}}{4}
\eee

\bee
\label{B15}
\Rightarrow\qquad \hbar_{eff}\,=\,\hbar\,\sqrt{1\,+\,\frac{\theta_{12}\eta_{12}}{\hbar^2}}\,\,.
\eee


\ni which is the NC correction to $\hbar$.

It is easy to see from (\ref{B15}) that 
\bee
\label{4.8}
\hbar_{eff}\,=\,\hbar\,\sqrt{1\,+\,\frac{\theta\eta}{\hbar^2}} \approx \hbar\(1\,+\,\frac{\theta\eta}{2\hbar^2}\)\,\,,
\eee
where we have expanded the square root term in (\ref{B15}) and considered only first order terms in $\theta\eta$, for obvious reasons.  

In the context of the Generalized Uncertainty Principle (GUP), in \cite{ghosh} for example, where noncommutativity was considered too, we can define an $\hbar_{eff}$ with the NC parameter within, which brought in a minimum length scale.  Hence, we could imagine if (\ref{4.8}) has an analogous effect and if we could construct, with Eq. (\ref{4.8}), a kind of GUP algebra.  But these considerations are out of the scope of this paper.

By using a simple calculation \cite{bbdp} where we assume that $\theta \eta / \hbar^2 \ll 1$ we obtain that
\bee
\label{4.9}
F_{NC}\,=\,\frac{GMm}{r^2}\,\(1\,+\,\frac{\theta\eta}{2\hbar^2}\)\,\,.
\eee
Hence,
\bee
\label{4.10}
G_{NC}\,=\,G\,\(1\,+\,\frac{\theta\eta}{2\hbar^2}\)\,\,.
\eee
From (\ref{S}) we can write that
\bee
\label{4.11}
G_{NE}\,=\,G\frac{5-3q}{2}\,=\,G\[1\,+\,\frac 32 (1-q) \]\,\,.
\eee

Back to Eq. (\ref{S}) we saw that when $q=1$ we have $G_{NE}=G$.  Now we will make the inverse analysis.  Our mathematical (let us call this way) motivation to use noncommutativity here is to obtain a new bound for Tsallis parameter.  So, the question we have to make now, differently from the one made in Eq. (\ref{S}) is, what is the value of $q$ when we analyze the Tsallis formalism at Planck scale (with noncommutativity)?  In other words, what is the value of $q$ such that $G_{NE}=G_{NC}$?  This is the inverse of what happened when we discussed the limit $q=1$ in Eq. (\ref{S}).

Having said that, comparing Eqs. (\ref{4.10}) and (\ref{4.11}), we have that
\bee
\label{4.13}
G_{NE} = G_{NC}\,\,,
\eee
we must write
\bee
\label{4.12}
\frac{\theta\eta}{2\hbar^2}\,=\,\frac 32 (1 - q)\,\,,
\eee
which is the introduction of the NC parameter into Tsallis formalism via $q$.  Hence, the $q$-parameter makes the connection between both theories, since the $|q-1|$ bound can be calculated.  From (\ref{4.12}), we can conclude that it is natural to expect very low values for $|q-1|$ due to the Planck scale properties of the NC objects involved in Eq. (\ref{4.12}).

The introduction of $G_{NE}$ means the introduction of a fractal geometry in the original system.  As the fractal geometry is connected to NC geometry, concerning fractal objects, the principle of self-similarity denies the notion of a simple geometrical point like the idea of differentiability.

Consequently from (\ref{4.12}) we can write that
\bee
\label{4.14}
1\,-\,q\,=\,\frac{\theta\eta}{3\hbar^2}\,\,.
\eee


Hence, due to the experimental results in  \cite{scwga} we obtain that  \cite{bbdp}
\baa
\frac{\theta \eta}{\hbar^2}\leq\,O(1)\cdot 10^{-13},
\eaa
and therefore we have that
\bee
\label{4.17}
|q-1|\leq O(1)\cdot 10^{-13},
\eee
which establish a new bound for the NE parameter.  This value is $10^{9}$ times below the bounds obtained 
in  \cite{TBL} and  \cite{TT}.   This is the main result of our work.

In \cite{DJ} the authors promoted the equivalence between NC and Tsallis formulations of magnetic susceptibilities which were calculated using only the first order term $(1-q)$ in a specific expansion.  Notice that from (\ref{4.10}), (\ref{4.11}) and (\ref{4.12}) we have no expansion in $(1-q)$ although we have a well known expansion of (\ref{4.8}).  The difference is that the reason to use only the first order term in \cite{DJ} is that the value $(1-q)$ is well known in the literature as being small.  In (\ref{4.12}) we do not need to know previously the value of $(1-q)$ to perform any kind of expansion.

However, it is important to say, that the equivalence obtained in (\ref{4.13}) does not mean necessarily, that it can be applied to any other physical systems.  As in \cite{DJ} we also believe that it can be an object for additional studies.

From Eq. (\ref{2c}) we see clearly that when $q \rightarrow 1$ we recover Boltzmann theory.  From Eq. (\ref{4.14}) we noticed that in this limit the term $\theta \eta / 3 \hbar^2$ goes to zero and the NC contribution goes to zero too.  Hence, as expected, to recover Boltzmann theory and to consider a NC space are completely different scenarios.  This fact is not true concerning Tsallis theory since we have the bound in Eq. (\ref{4.17}) obeyed.

\section{Friedmann Equations Using Tsallis Statistics}
\renewcommand{\theequation}{5.\arabic{equation}}
\setcounter{equation}{0}
\label{FET}

In order to make this paper self-contained, in the first part of this section we will review the main steps of the obtainment of the Friedmann equations through the entropic force theory.  
This calculation will be followed by the obtainment of the same equation through the NE Tsallis statistics.
Our objective is to obtain the same expression for the gravitational constant obtained in the last section.  We think that this result confirms the relation between both constants.  After that we will discuss some nonextensivity cosmological effects.

In what follows we will review the calculation performed in  \cite{cco} where the Friedmann equations, which rules the dynamical evolution of the Friedmann-Robertson-Walker Universe, were obtained through the entropic force theory together with the equipartition law of energy.  The authors also used the concept of Unruh temperature combined with Verlinde's work.

\subsection{Friedmann Equations from Entropic Force: a quick review}

Let us consider the FRW metric given by

\bee
\label{AA}
ds^2\,-\,dt^2\,+\,a^2(t)\,(dr^2\,+\,r^2 d\Omega^2 ),
\eee

\ni where $a(t)$ is the scale factor of the Universe.  Following Verlinde's point of view  \cite{verlinde}, we will consider a compact spatial region ${\cal S}$ with a compact boundary $\partial {\cal S}$, which is a sphere with physical radius $\tilde{r}\,=\,a\,r$.  In this framework, this compact boundary plays the role of the holographic screen.  Concerning relativity, we can write that $E=Mc^2$, where $M$ represents the mass that would emerge in the compact spatial region ${\cal S}$ surrounded by the boundary screen $\partial {\cal S}$.

We will assume that the FRW Universe is a perfect fluid with stress-energy tensor given by

\bee
\label{DD}
T_{\mu\nu}\,=\,(\rho\,+\,p)\,u_{\mu} u_{\nu}\,+\,p g_{\mu\nu},
\eee

\ni where the total mass is not conserved anymore.  We can consider that the change in the total mass is equal to the work given by the pressure so that $dM=-pdV$.  Consequently we have the continuity equation

\bee
\label{EE}
\dot{\rho}\,+\,3 H (\rho\,+\,p)\,=\,0\quad \mbox{where} \quad H\,=\,\frac{\dot{a}}{a} \quad (\mbox{Hubble parameter})\,.
\eee
The total mass inside the spatial region $\Lambda$ is given by

\bee
\label{FF}
M\,=\,\int_{\Lambda}\,dV\, T_{\mu\nu} u^{\mu} u^{\nu},
\eee

\ni where $T_{\mu\nu} u^{\mu} u^{\nu}$ is the energy density measured by a comoving observer.  A comoving observer at the place of the screen $r$ can measure the acceleration as

\bee
\label{GG}
a_r\,=\,-\,\frac{d^2 \tilde{r}}{dt^2}\,=\,-\,\ddot{a}{r}\,\,.
\eee

\ni This acceleration is caused by the matter in the spatial region enclosed by the boundary $\partial {\cal S}$

The Unruh formula (\ref{un}) can be written as

\bee
\label{HH}
T\,=\,\frac{1}{2\pi k_B c}\,\hbar a_r \quad \Rightarrow k_B T\,=\,-\frac{1}{2\pi c}\,\hbar \ddot{a} r\,\,,
\eee

\ni where we can realize the relation between the acceleration and the temperature.  Substituting  Eqs. (\ref{bits}), (\ref{HH}) and the relativistic energy in Eq. (\ref{eq})  we obtain that

\bee
\label{II}
\ddot{a}\,=\,-\frac{4\pi G}{3}\,\rho a \, ,
\eee
\ni which determines the evolution of $a(t)$, i.e., it is the dynamical equation for Newtonian cosmology \cite{cco}.  Eq. (\ref{II}), together with the equation for the evolution of the energy density, coincide with the equations for dust in General Relativity.  We can also obtain Eq. (\ref{II}) from the Newtonian gravity law but here it was obtained via holographic principle.

Let us define the active gravitational mass ${\cal M}$, different from the total mass in the spatial region ${\cal S}$.  It is the well known Tolman-Komar mass, defined by

\baa
\label{JJ}
{\cal M}\,&=&\,2\,\int_\Lambda dV \Big( T_{\mu\nu}\,-\frac 12 T g_{\mu\nu} \Big) u^{\mu} u^{\nu} \nonumber \\
&=&\frac{4\pi}{3}\,a^3\,r^3\,\Big( \rho\,+\,\frac{3p}{c^2} \Big)\, .
\eaa


\ni Using Eqs. (2.1) and (2.3), where $M$ will be replaced by the active gravitational mass, and Eqs. (\ref{JJ}) 
and  (\ref{un}) we obtain

\bee
\label{LL}
\frac{\ddot{a}}{a}\,=\,-\,\frac{4\pi G}{3}\,(\rho\,+\,3p).
\eee

\ni This is the well known acceleration equation for the dynamical evolution of the FRW Universe.  Using the continuity equation (\ref{EE}) and after an integration we can obtain the Friedmann equations for the FRW Universe given by

\bee
\label{MM}
H^2\,+\,\frac{k}{a^2}\,=\,\frac{8\pi G}{3}\,\rho ,
\eee

\ni where $k$ is an integration constant which can be realized as the spatial curvature in the region ${\cal S}$ in the Einstein theory of general relativity.  The possible values of $k$ defines the geometry of the FRW Universe and they are well known $k=1$ (closed), $0$ (flat) and $-1$ (open).
Physically, Eq. (\ref{MM}) can be considered as the energy conservation equation for a rocket launched from the surface of the Earth with unit mass and speed $\dot{a}$ \cite{mukhanov}.  The normalization of $a$ has no invariant meaning in Newtonian and consequently it can be rescaled by an arbitrary factor.

The reader can see more details in  \cite{verlinde} about the extension of the above calculation for the case of extra dimensions, i.e., $d\geq 4$.  For example, the number of bits on the screen is now given by

\bee
\label{A1}
N\,=\,\frac 12\,\frac{d-2}{d-3}\,\frac{Ac^3}{G\hbar} ,
\eee

\ni and the continuity equation is

\bee
\label{A2}
\dot{\rho}\,+\,(d-1) H (\rho\,+\,p)\,=\,0.
\eee

\ni  The active mass is

\bee
\label{A3}
{\cal M}\,=\,\frac{d-2}{d-3}\,\int_\Lambda\,dV\, \Big( T_{\mu\nu}\,-\,\frac{1}{d-2}\,T g_{\mu\nu} \Big)\,u^{\mu} u^{\nu},
\eee

\ni  and the acceleration given in (\ref{LL}) will be written as

\bee
\label{A4}
\frac{\ddot{a}}{a}\,=\,-\,\frac{8\pi G}{(d-1)(d-2)}\, [(d-3)\rho\,+\,(d-1) p].
\eee

\ni   Finally, the Friedmann equation for the FRW Universe can be represented in $d$ dimensions by

\bee
\label{A5}
H^2\,+\,\frac{k}{a^2}\,=\,\frac{16\pi G}{(d-1)(d-2)}.
\eee
This is the holographic principle derivation of the Friedmann equations of the FRW Universe.  We basically used only the entropic ideas derived by Verlinde through the equipartition law of energy.  The main argument obviously is that gravity appears as an entropic force.

In the next subsection, as explained before, we will show the resulting equations originating from the NE ideas developed by Tsallis.

\subsection{The Modified Friedmann Equations through Tsallis theory}

Using Eq. (\ref{ge}), which describes the NE concept of equipartition law of energy 
and by following the same calculation developed above we can obtain the NE acceleration equation for the dynamical evolution of the FRW Universe

\bee
\label{A8}
\frac{\ddot{a}}{a}\,=\,-(5-3q)\,\frac{2\pi G}{3}\,(\rho\,+\,3p),
\eee

\ni and in a direct calculation we can write the (NE) Friedmann equation

\bee
\label{A9}
H^2\,+\,\frac{k}{a^2}\,=\,\frac{4(5-3q)\pi G}{3}\rho a^2,
\eee

\ni which is easy to compare with Eq. (\ref{MM}) and consequently we obtain Eq. (\ref{S})
which is the same result obtained in the section \ref{TTS}.  So, we consider this outcome as a confirmation that the NE equipartition law of energy introduces the NE parameter in the gravitational constant.  We can understand Eq. (\ref{S}) as a numerical map between both constants in such a way to produce a direct transformation from one concept to another.  In other words we can say that a simple substitution determined by Eq. (\ref{S}) produces the introduction of Tsallis concept into any theory.
For example, it is easy to obtain the dimensional generalization of the Friedmann equation in $d$ dimension.  This equation can be written as 

\bee
\label{A11}
H^2\,+\,\frac{k}{a^2}\,=\,\frac{8\pi G_{NE}}{(d-1)(d-2)}\,\rho,
\eee

\ni and $G_{NE}$ is given by Eq. (\ref{S}).  This result shows us that we can introduce the nonextensive principle into cosmological objects like Friedmann equations.  As we have discussed before, NC theory and TT are both connected by the fractal nature of their geometries.  Moreover, as the NC fractal nature is carried by the NC parameter, we could see from Eq. (\ref{4.12}) that its connection with the $q$-parameter would be expected.  We will see the application of these ideas at the cosmological level.

\subsection{Cosmological parameters analysis}

In this section we will analyze some cosmological parameters in the light of Tsallis nonextensive statistics.  Let us begin with the density parameter given by

\bee
\label{Aa}
\Omega_M \,=\,\frac{\rho}{\rho_c}\,\,,
\eee

\ni where $\rho_c$ is the critical density that, for a flat Universe, $k=0$, is written as

\bee
\label{Bb}
H^2\,-\,\frac 83\,\pi G \rho_c\,=\,0 \quad \Longrightarrow \quad \rho_c\,=\,\frac{3H^2}{8\pi G}\,\,,
\eee

\ni From Eq. (\ref{S}) we have that

\bee
\label{Cc}
\rho_c\,=\,\frac{3H^2 (5\,-\,3q)}{16\pi G_{NE}}
\eee

\ni and if $q=5/3$ we have that $\rho_c=0$.  We have that for $\rho_c=0$, the Universe is closed.  As we said before, if $q=1\,(\Rightarrow G_{NE}=G)$, we recover Boltzmann-Gibbs and, from (\ref{Cc})

\bee
\label{Dd}
\rho_c \,=\,\rho_c^{BG}\,=\,\frac{3H^2}{8\pi G}\,\,,
\eee

\ni which coincides with Eq. (\ref{Bb}).

Back to the density parameter, let us substitute Eq. (\ref{Cc}) into Eq. (\ref{Aa}) and we have that

\bee
\label{DdDd}
\Omega_M \,=\,\frac{16\pi \rho G_{NE}}{3H(5-3q)}\,\,,
\eee

\ni which, for a flat Universe is equal to one, i.e., $\Omega_M\,=\,1$, so

\bee
\label{Ee}
\frac{16\pi \rho G_{NE}}{3H(5-3q)}\,=\,1 \quad \Longrightarrow \quad q\,=\,\frac 13\,\Big[5\,-\,\frac{16\pi G_{NE}}{3H}\,\rho \Big]
\eee

\ni and we have a direct relation between $q$, the matter density and the Hubble parameter for a flat Universe.  In Eq. (\ref{DdDd}), if $q=5/3$ we have that $\Omega_M \longrightarrow \infty$, which characterizes a closed Universe.  
For $q\,<\,5/3\:\Rightarrow\:\Omega_M$ grows in an inverse relation with $q$.  We can conclude that $q$ could be intrinsically related to the geometry of the Universe.

However, with a cosmological constant, the critical density is no longer a well-defined quantity.  So, let us describe three well known quantities, the matter density, the energy density and the curvature density of the Universe.  They are given respectively by

\bee
\label{Ff}
\Omega_M\,=\,\frac{\rho}{\rho_c}\,=\,\frac{16\pi G_{NE}\rho_0}{3H_0(5-3q)}\,\,,
\eee

\bee
\label{Gg}
\!\!\!\!\!\!\!\!\!\!\!\!\!\!\!\!\!\!\!\!\!\!\!\!\!\!\!\!\!\!\!\!\!\!\!\!\!\!\!\Omega_{\Lambda}\,=\,\frac{\Lambda}{3H_0^2}\,\,
\eee
and
\bee
\label{Hh}
\!\!\!\!\!\!\!\!\!\!\!\!\!\!\!\!\!\!\!\!\!\!\!\!\!\!\!\!\!\!\!\!\!\!\Omega_k\,=\,-\frac{k}{a_0^2\,H_0^2}
\eee

\ni where the zero indice denotes ``current value."

We also have the constraint

\bee
\label{Ii}
\Omega_M\,+\,\Omega_{\Lambda}\,+\,\Omega_k\,=\,1\,\,.
\eee

\ni To discuss, for example, this constraint with Tsallis NE considerations we use a flat Universe so that $k=0 \Rightarrow \Omega_k = 0$, and from Eq. (\ref{Ii})

\bee
\label{Jj}
\Omega_{\Lambda}\,=\,1\,-\,\Omega_M \quad \Longrightarrow \quad \Omega_{\Lambda}\,=\,1\,-\,\frac{16\pi G_{NE} \rho_0}{3(5-3q)H_0}
\eee

\ni and, as we have the constraint $q\,<\,5/3$, we have, from Eq. (\ref{Gg}) and (\ref{Jj}) that, for a flat Universe, the cosmological constant will be given by,

\bee
\label{Kk}
\Lambda\,=\,3H_0\,-\,\frac{16\pi G_{NE}\rho_0}{5-3q}\,\,,
\eee

\ni which shows a nonlinear relation between the cosmological constant and the $q$-parameter.

To end this section, we saw that the consideration of the NE principle can lead us to perturb some cosmological parameters and the Universe geometry.  As we could understand, this perturbation revealed itself in a way such that the NE parameter can be considered as an important object in cosmology.  We can ask further in what way can the q-bound obtained here can be present or perturb in the current cosmological data.

\section{Conclusions}
\label{Cs}

The main motivation to connect the theory developed by Tsallis and the NC geometry is the fact that both can be related to a fractal characteristic of the background space.  After this fact be well understood, we can ask for which values of the $q$-parameter we can introduce the NC parameter in TT.  We demonstrated precisely here that this calculation establish a new bound for the NE parameter in the literature.  Besides, we have demonstrated that the relation between $G_{NE}$ and $G$ can also be obtained through Friedmann equation.  Finally, we analyzed the influence that the NE principle have in other cosmological parameters like the energy density, cosmological constant, etc..

We have applied the NE equipartition law of energy in Verlinde's formalism of gravity.  As a result we have obtained an effective gravitational constant that incorporates  effects of the NE statistics developed by Tsallis. On the other side, we have used the NC theory to predict an effective gravitational constant.

If we consider both gravitational (NC and NE) constants equal, a new bound for the Tsallis $q$-parameter can be derived. The result of the protocol used to obtain a NE Friedmann equations is simply to write the usual gravitational constant as  a function of the NE effective gravitational constant.

Facing these results we can ask ourselves if this connection between noncommutativity and Tsallis approach can affect other related subjects.  For example, it is well known that a correspondence like AdS/CFT is a manifestation of the UR/UV connection and that it has a quite closed relationship with string theory.  At the same time we also know that a string theory embedded in a magnetic background presents a NC algebra.  Since we saw in this work the connection between both NC and Tsallis formalism, we can think if we can obtain a connection between the AdS/CFT correspondence and Tsallis theory.
Also, we can imagine if the holographic screen can be formulated as a de Sitter (dS) surface.  The next step would be to study a test particle in a dS holographic screen submitted to an entropic force.  This is an ongoing research.

In \cite{mateus} three of us analyzed the Universe as a perfect fluid through its Lagrangian and Hamiltonian formulations with the cosmological constant.  Using the symplectic embedding technique, it was introduced the NC parameter and its physical consequences were precisely depicted.  We believe that we now have a path to an alternative understanding of the main features of the Universe such as its homogeneity and isotropy and the constraints of the cosmological constant via Tsallis formalism, some of these concepts were exemplified in the last section, where we hope these concepts were clarified a little bit more.

What is also new here is to establish the connection between both NC and NE Tsallis theories through Verlinde's idea.  Although we have explored the results obtained in \cite{bbdp}, our main objective here was to determine a new bounds for $q$.
As another perspective it can be analyzed the compatibility of MOND with NE Friedmann equations.

\section{Acknowledgments}

\bigskip
\bigskip

The authors would like to thank the anonymous referees that, with their requests, turned this paper into a better one.
EMCA and WO would like to thank CNPq, Conselho Nacional de Desenvolvimento Cient\' ifico e Tecnol\'ogico, a Brazilian research support agency, for partial financial support.

\end{document}